\documentclass[aps,groupedaddress,showpacs,showkeys]{revtex4}
\begin{document}
\def\Barcelo{Barcel\'o}
\title{Analog gravity from field theory normal modes?}
\author{Carlos \Barcelo}
\email[]{carlos@hbar.wustl.edu}
\homepage[]{http://www.physics.wustl.edu/~carlos}
\thanks{Supported by the Spanish MEC}
\affiliation{Physics Department, Washington University, 
Saint Louis, MO 63130-4899, USA}
\author{Stefano Liberati}
\email[]{liberati@physics.umd.edu}
\homepage[]{http://www2.physics.umd.edu/~liberati}
\thanks{Supported by the US NSF}
\affiliation{Physics Department, University of Maryland, 
College Park, MD 20742-4111, USA}
\author{Matt Visser}
\email[]{visser@kiwi.wustl.edu}
\homepage[]{http://www.physics.wustl.edu/~visser}
\thanks{Supported by the US DOE}
\affiliation{Physics Department, Washington University, 
Saint Louis MO 63130-4899, USA}
\date{2 April 2001;  \LaTeX-ed \today}
\bigskip
\begin{abstract}
\bigskip

We demonstrate that the emergence of a curved spacetime ``effective
Lorentzian geometry'' is a common and generic result of linearizing a
classical scalar field theory around some non-trivial background
configuration.  This investigation is motivated by considering the
large number of ``analog models'' of general relativity that have
recently been developed based on condensed matter physics, and asking
whether there is something more fundamental going on. Indeed,
linearization of a classical field theory (that is, a field theoretic
``normal mode analysis'') results in fluctuations whose propagation is
governed by a Lorentzian-signature curved spacetime ``effective
metric''. In the simple situation considered in this paper, (a single
classical scalar field), this procedure results in a unique effective
metric, which is quite sufficient for simulating kinematic aspects of
general relativity (up to and including Hawking radiation). Upon
quantizing the linearized fluctuations around this background
geometry, the one-loop effective action is guaranteed to contain a
term proportional to the Einstein--Hilbert action of general
relativity, suggesting that while classical physics is responsible for
generating an ``effective geometry'', quantum physics can be argued to
induce an ``effective dynamics''. The situation is strongly
reminiscent of, though not identical to, Sakharov's ``induced
gravity'' scenario, and suggests that Einstein gravity is an emergent
low-energy long-distance phenomenon that is insensitive to the details
of the high-energy short-distance physics. (We mean this in the same
sense that hydrodynamics is a long-distance emergent phenomenon, many
of whose predictions are insensitive to the short-distance cutoff
implicit in molecular dynamics.)

\end{abstract}
\maketitle

\def\half{{1\over2}}
\def\L{{\mathcal L}}
\def\S{{\mathcal S}}
\def\d{{\mathrm{d}}}
\def\etal{{\emph{et al}}}
\def\det{{\mathrm{det}}}
\def\tr{{\mathrm{tr}}}
\def\ie{{\emph{i.e.}}}
\def\aka{{\emph{aka}}}
\def\Choose#1#2{{#1 \choose #2}}
\def\Eotvos{{E\"otv\"os}}
\def\background{{\mathrm{b}}}
\def\quantum{{\mathrm{quantum}}}
\def\HRULE{{\bigskip\hrule\bigskip}}

\section{Introduction}

The idea of building analog models of, and possibly for, general
relativity is currently attracting considerable
attention~\cite{Workshop}. Because of the extreme difficulty (and
inadvisability) of working with intense gravitational fields in a
laboratory setting, interest has now turned to investigating the
possibility of {\emph{simulating}} aspects of general relativity ---
though it is not {\emph{a priori}} expected that all features of
Einstein gravity can successfully be carried over to the condensed
matter realm.  Numerous rather different physical systems have now
been seen to be useful for developing analog models of general
relativity:
\begin{enumerate}
\item
Dielectric media: A refractive index can be reinterpreted as an
effective metric, the Gordon metric. (Gordon~\cite{Gordon},
Skrotskii~\cite{Skrotskii}, Balazs~\cite{Balazs},
Plebanski~\cite{Plebanski}, de Felice~\cite{deFelice}, and many
others.)
\item
Acoustics in flowing fluids: Acoustic black holes, {\aka} ``dumb
holes''. (Unruh~\cite{Unruh}, Jacobson~\cite{Jacobson},
Visser~\cite{Visser}, Liberati \etal~\cite{Liberati}, and many
others.)
\item 
Phase perturbations in Bose--Einstein condensates: Formally similar to
acoustic perturbations, and analyzed using the nonlinear Schrodinger
equation (Gross--Pitaevskii equation) and Landau--Ginzburg Lagrangian;
typical sound speeds are centimetres per second to millimetres per
second. (Garay {\etal}~\cite{Garay}, {\Barcelo}~\cite{Barcelo}
{\etal}.)
\item
High-refractive-index dielectric fluids (``slow light''): In
dielectric fluids with an extremely high group refractive index it is
experimentally possible to slow lightspeed to centimetres per second
or less. (Leonhardt--Piwnicki~\cite{Leonhardt}, Hau
{\etal}~\cite{Hau}, Visser~\cite{Visser2}, and others.)
\item
Quasi-particle excitations: Fermionic or bosonic quasi-particles in a
heterogeneous superfluid environment. (Volovik~\cite{Volovik},
Kopnin--Volovik~\cite{Kopnin},
Jacobson--Volovik~\cite{Jacobson-Volovik}, and
Fischer~\cite{Fischer}.)
\item
Nonlinear electrodynamics: If the permittivity and permeability
themselves depend on the background electromagnetic field, photon
propagation can often be recast in terms of an effective metric.
(Plebanski~\cite{Plebanski2}, Dittrich--Gies~\cite{Dittrich}, Novello
{\etal}~\cite{Novello}.)
\item
Linear electrodynamics: If you do not take the spacetime metric itself
as being primitive, but instead view the linear constitutive
relationships of electromagnetism as the fundamental objects, one can
nevertheless reconstruct the metric from first principles. (Hehl,
Obukhov, and Rubilar~\cite{Obukhov-Hehl,Hehl,Obukhov}.)
\item
Scharnhorst effect: Anomalous photon propagation in the Casimir vacuum
can be interpreted in terms of an effective metric.
(Scharnhorst~\cite{Scharnhorst}, Barton~\cite{Barton}, Liberati
{\etal}~\cite{Liberati2}, and many others.)
\item
Thermal vacuum: Anomalous photon propagation in QED at nonzero
temperature can be interpreted in terms of an effective metric.
(Gies~\cite{Gies}.)
\item
``Solid state'' black holes. (Reznik~\cite{Reznik}, Corley and
Jacobson~\cite{Corley}, and others.)
\item
Astrophysical fluid flows: Bondi--Hoyle accretion and the Parker wind
[coronal outflow] both provide physical examples where an effective
acoustic metric is useful, and where there is good observational
evidence that acoustic horizons form in nature. (Bondi~\cite{Bondi},
Parker~\cite{Parker-wind}, Moncrief~\cite{Moncrief},
Matarrese~\cite{Matarrese}, and many others.)
\item
Other condensed-matter approaches that don't quite fit into the above
classification~\cite{Hu,Chapline}.
\end{enumerate}
A literature search as of April 2001 finds well over a hundred
scientific articles devoted to one or another aspect of analog gravity
and effective metric techniques. The sheer number of different
physical situations lending themselves to an ``effective metric''
description strongly suggests that there is something deep and
fundamental going on. Typically these are models {\em of} general
relativity, in the sense that they provide an effective metric and so
generate the basic kinematical background in which general relativity
resides; in the absence of any dynamics for that effective metric we
cannot really speak about these systems as models {\em for} general
relativity. However, as we will discuss more fully bellow, quantum
effects in these analog models might provide of some sort of dynamics
resembling general relativity.

On a related front, (and we'll see the connection soon enough), the
particle physics and relativity communities have also seen Lorentz
symmetry emerge as a low-energy approximate symmetry in several
physical situations:
\begin{enumerate}
\item
As an infra-red fixed point of the renormalization group in certain
non-Lorentz invariant quantum field theories; (Nielsen
{\etal}~\cite{Nielsen}).
\item
As a low-momentum approximation to acoustic propagation in the
presence of viscosity; (Visser~\cite{Visser}).
\item
As a low-momentum approximation to quasi-particle propagation governed
by the Bogolubov dispersion relation; ({\Barcelo} {\etal}~\cite{Barcelo}).
\item
In certain other quasi-particle dispersion relations
(Volovik~\cite{Volovik-disp}).
\end{enumerate}

On a third front, the last few years have seen an increasing number of
indications that Einstein gravity (and even quantum field theory) may
not be as ``fundamental'' as was once supposed:
\begin{enumerate}
\item
Induced gravity: In ``induced-gravity'' models a la
Sakharov~\cite{Sakharov} the dynamics of gravity is an emergent
low-energy phenomenon that is {\em not} fundamental physics. In those
models the dynamics of gravity (the approximate Einstein equations) is
a consequence of the quantum fluctuations of the other fields in the
theory. (In induced gravity models gravitation is not fundamental in
exactly the same sense that phonons are not fundamental: phonons are
collective excitations of condensed matter systems.  Phonons are not
fundamental particles in the sense of, say, photons. But this should
not stop you from quantizing the phonon field as long as you realise
you should not take phonons seriously at arbitrarily high momenta.)
\item 
Effective field theory for gravity: Donoghue~\cite{Donoghue} has
strongly argued that quantum gravity itself should simply be viewed as
an effective field theory, in the same sense that the Fermi theory of
the weak interactions is an effective field theory --- it still makes
sense to quantize in terms of gravitons~\cite{Feynman}, but the
high-energy physics is likely to be rather different from what could
be guessed based only on observing low-energy excitations, and you
should not necessarily take the gravitons seriously at arbitrarily
high momenta.
\item 
Effective field theory in general: Indeed, even strictly within the
confines of particle physics, attitudes towards effective theories
seem to be changing --- they are now much more likely to be used, at
least as computational tools. As long as one has a clear understanding
of when to stop taking them seriously effective theories are perfectly
good physics even if they are not
``fundamental''~\cite{Weinberg1,Weinberg2}.
\end{enumerate}

These various observations led us to suspect the existence of a
general pattern: That the occurrence of something like an approximate
Lorentz symmetry, and something like an approximate non-trivial
``effective metric'' might be an inescapable general consequence of
classical and quantum field theories viewed as dynamical systems. In
this article we take some important steps in this direction, and
indicate the issues that still must be tackled.

Remember that for mechanical systems with a finite number of degrees
of freedom small oscillations can always be resolved into normal
modes: a finite collection of uncoupled harmonic oscillators.  For a
classical field theory you would also expect similar behaviour: small
deviations from a background solution of the field equations will be
resolved into travelling waves; then these travelling waves can be
viewed as an infinite collection of harmonic oscillators, or a finite
number if the field theory is truncated in the infra-red and
ultra-violet, to which you can then apply a normal mode analysis.  The
physically interesting question is whether this normal mode analysis
for field theories can then be reinterpreted in a ``geometrically
clean'' way in terms of some ``effective metric'' and ``effective
geometry''.  We shall show that whenever we are dealing with a single
scalar field the answer is definitely yes: Linearization of any
Lagrangian-based dynamics, or linearization of any hyperbolic
second-order PDE, will automatically lead to an effective Lorentzian
geometry that governs the propagation of the fluctuations. [The
general situation (multiple scalar fields, or a multi-component vector
or tensor) is quite algebraically messy --- details of that situation
will be deferred for now.]

Once we have developed the notion of a derived ``effective metric''
based on a linearization procedure, we can certainly consider the
effect of quantizing the linearized fluctuations. At one loop the
quantum effective action will contain a term proportional to the
Einstein--Hilbert action --- this is, in modern language, a key
portion of Sakharov's ``induced gravity'' idea~\cite{Sakharov}. In the
closing segment of the present article we argue that the occurrence of
not just an ``effective metric'', but also an ``effective
geometrodynamics'' closely related to Einstein gravity is a largely
unavoidable feature of the linearization and quantization process.

\section{Lagrangian analysis}
\label{S:lagrangian}

Suppose we have a single scalar field $\phi$ whose dynamics is
governed by some first-order Lagrangian $\L(\partial_\mu\phi, \phi)$.
(By ``first-order'' we mean that the Lagrangian is some arbitrary
function of the field and its first derivatives.) We want to consider
linearized fluctuations around some background solution $\phi_0(t,\vec
x)$ of the equations of motion, and to this end we write
\begin{equation}
\phi(t,\vec x) = \phi_0(t,\vec x) + \epsilon  \phi_1(t,\vec x) + 
 {\epsilon^2\over2}  \phi_2(t,\vec x) + O(\epsilon^3).
\end{equation}
Now use this to expand the Lagrangian around the classical solution
$\phi_0(t,\vec x)$:
\begin{eqnarray}
{\L}(\partial_\mu \phi,\phi) 
&=& 
{\L}(\partial_\mu \phi_0,\phi_0)
+
\epsilon \left[ 
{\partial \L\over\partial(\partial_\mu \phi)} \; \partial_\mu \phi_1
+
{\partial \L\over\partial\phi} \; \phi_1
\right]
+
{\epsilon^2\over2} \left[ 
{\partial \L\over\partial(\partial_\mu \phi)} \; \partial_\mu \phi_2
+
{\partial \L\over\partial\phi} \;  \phi_2
\right]
\nonumber
\\
&+&
{\epsilon^2\over2} \Bigg[ 
{\partial^2 \L\over\partial(\partial_\mu \phi) \; \partial(\partial_\nu \phi)} 
\; \partial_\mu \phi_1 \; \partial_\nu \phi_1
+
2 {\partial^2 \L\over \partial(\partial_\mu \phi)\; \partial \phi} 
\;\partial_\mu \phi_1 \; \phi_1
+
{\partial^2 \L\over \partial\phi\; \partial\phi} 
\; \phi_1 \; \phi_1
\Bigg]
+
O(\epsilon^3).
\end{eqnarray}
It is particularly useful to consider the action
\begin{equation}
\label{E:bare-action}
S[\phi] = \int \d^{d+1} x \; \L(\partial_\mu\phi,\phi),
\end{equation}
since doing so allows us to integrate by parts. (Note that the
Lagrangian $\L$ is taken to be a tensor density, not a scalar.)  We
can now use the Euler--Lagrange equations for the background field
\begin{equation}
\partial_\mu \left( {\partial \L \over \partial(\partial_\mu \phi)} \right) 
- {\partial \L\over\partial \phi} = 0,
\end{equation}
to discard the linear terms (remember we are linearizing around a
solution of the equations of motion) and so we get
\begin{eqnarray}
S[\phi] &=& S[\phi_0] 
+
{\epsilon^2\over2}
\int \d^{d+1} x \Bigg[
\left\{
  {\partial^2 \L\over
   \partial(\partial_\mu \phi) \; \partial(\partial_\nu \phi)} 
   \right\} 
\; \partial_\mu \phi_1 
\; \partial_\nu \phi_1 
+
\left(
{\partial^2 \L\over \partial\phi\; \partial \phi} - 
\partial_\mu \left\{{\partial^2 \L\over \partial(\partial_\mu \phi)\; 
\partial \phi} \right\}
\right) 
\;
\phi_1 \; \phi_1
\Bigg]
+
O(\epsilon^3).
\end{eqnarray}
Having set things up this way, the equation of motion for the
linearized fluctuation is now easily read off as
\begin{eqnarray}
&&  \partial_\mu \left(\left\{
{\partial^2 \L\over\partial(\partial_\mu \phi) \; \partial(\partial_\nu \phi)} 
   \right\}
 \partial_\nu \phi_1 \right)
- \left(
{\partial^2 \L\over \partial\phi\; \partial \phi} - 
\partial_\mu \left\{{\partial^2 \L\over \partial(\partial_\mu \phi)\; 
\partial \phi} \right\}
\right)
\phi_1
= 0. 
\end{eqnarray}
This is a second-order differential equation with position-dependent
coefficients (these coefficients all being implicit functions of the
background field $\phi_0$). Following an analysis developed for
acoustic geometries (Unruh~\cite{Unruh}, Visser~\cite{Visser},
Liberati {\etal}~\cite{Liberati}, {\Barcelo} {\etal}~\cite{Barcelo}),
which we now apply to this much more general situation, the above can
be given a nice clean geometrical interpretation in terms of a
d'Alembertian wave equation --- provided we \emph{define} the
effective spacetime metric by
\begin{equation}
\sqrt{-g} \; g^{\mu\nu} \equiv  f^{\mu\nu} \equiv 
\left.\left\{
{\partial^2 \L\over\partial(\partial_\mu \phi) \; \partial(\partial_\nu \phi)} 
\right\}\right|_{\phi_0}.
\end{equation}
Suppressing the $\phi_0$ except when necessary for clarity, this
implies [in (d+1) dimensions, $d$ space dimensions plus $1$ time
dimension]
\begin{equation}
(-g)^{(d-1)/2} = -\det   
\left\{ 
{\partial^2 \L\over\partial(\partial_\mu \phi) \; \partial(\partial_\nu \phi)} 
\right\}.
\end{equation}
Therefore
\begin{equation}
 g^{\mu\nu}(\phi_0) =  
\left.
\left(
- \det   
\left\{
{\partial^2 \L\over\partial(\partial_\mu \phi) \; \partial(\partial_\nu \phi)} 
\right\}
\right)^{-1/(d-1)}
\right|_{\phi_0}
\;
\left.
\left\{
{\partial^2 \L\over\partial(\partial_\mu \phi) \; \partial(\partial_\nu \phi)} 
\right\}
\right|_{\phi_0}.
\end{equation}
And, taking the inverse
\begin{equation}
g_{\mu\nu}(\phi_0) =    
\left.
\left( - 
\det\left\{
{\partial^2 \L\over\partial(\partial_\mu \phi) \; \partial(\partial_\nu \phi)} 
\right\}
\right)^{1/(d-1)}
\right|_{\phi_0}
 \; \; 
\left.
\left\{
{\partial^2 \L\over\partial(\partial_\mu \phi) \; \partial(\partial_\nu \phi)} 
\right\}^{-1}
\right|_{\phi_0}.
\end{equation}
We can now write the equation of motion for the linearized
fluctuations in the geometrical form
\begin{equation}
\label{E:geometrical}
\left[\Delta(g(\phi_0)) - V(\phi_0)\right] \phi_1 = 0,
\label{laeone}
\end{equation}
where $\Delta$ is the d'Alembertian operator associated with the
effective metric $g(\phi_0)$, and $V(\phi_0)$ is the
background-field-dependent potential
\begin{eqnarray}
V (\phi_0)&=&
{1\over\sqrt{-g}} \;
\left(
{\partial^2 \L\over \partial\phi\; \partial \phi} - 
\partial_\mu 
\left\{
{\partial^2 \L\over \partial(\partial_\mu \phi)\; \partial \phi} 
\right\}
\right)
\\
&=&
 \left(
- \det   
\left\{
{\partial^2 \L\over\partial(\partial_\mu \phi) \; \partial(\partial_\nu \phi)} 
\right\}
\right)^{-1/(d-1)}
\;
\left(
{\partial^2 \L\over \partial\phi\; \partial \phi} - 
\partial_\mu 
\left\{
{\partial^2 \L\over \partial(\partial_\mu \phi)\; \partial \phi} 
\right\}
\right). 
\end{eqnarray}
Thus $V(\phi_0)$ is a true scalar (not a density).  Note that the
differential equation (\ref{E:geometrical}) is automatically formally
self-adjoint (with respect to the measure $\sqrt{-g} \; \d^{d+1}x$).

It is possible to modify the metric by a conformal factor --- doing so
preserves the causal properties of the Lorentzian geometry but destroys
formal self-adjointness. (Nevertheless, one may be willing to pay this
price if the payoff is high enough.) Specifically let us define
\begin{equation}
\tilde g_{\mu\nu} \equiv  \exp(-2\theta) \;g_{\mu\nu}.
\end{equation}
Then
\begin{equation}
\sqrt{-\tilde g} \; \tilde g^{\mu\nu} \; \exp[(d-1)\theta] 
\equiv  f^{\mu\nu} \equiv 
\left.\left\{
{\partial^2 \L\over\partial(\partial_\mu \phi) \; \partial(\partial_\nu \phi)} 
\right\}\right|_{\phi_0}.
\end{equation}
A brief computation now yields
\begin{equation}
\label{E:geometrical2}
\left[\Delta(\tilde g(\phi_0)) 
+(d-1)\nabla\theta\cdot\nabla 
- \tilde V(\phi_0)\right] \phi_1 = 0,
\label{E:laeone-2}
\end{equation}
where the inner product is defined in terms of $\tilde g$ and now
\begin{equation}
\tilde  V(\phi_0) \equiv  \exp(2\theta) \; V(\phi_0).
\end{equation}
This could be used, for instance, to simplify the potential term at
the cost of self-adjointness. Generically, one would wish to preserve
formal self adjointness even at the cost of a more complicated
potential term, but we emphasize that at the level of causality there
is no strong reason for making a fixed choice.

It is important to realise just how general the result is (and where
the limitations are): it works for {\em any} Lagrangian depending only
on a single scalar field and its first derivatives. The linearized PDE
will be {\emph{hyperbolic}} (and so the linearized equations will have
wave-like solutions) if and only if the effective metric $g_{\mu\nu}$
has Lorentzian signature $\pm[-,+^d]$. Observe that if the
Lagrangian contains nontrivial second derivatives you should not be
too surprised to see terms beyond the d'Alembertian showing up in the
linearized equations of motion. Specific examples of this in special
cases are already known: for example this happens in the acoustic
geometry when you add viscosity (Visser~\cite{Visser}; not really a
Lagrangian system but the general idea is the same), or in the quantum
geometry of the Bose--Einstein condensate if you keep terms arising
from the quantum potential ({\Barcelo} {\etal}~\cite{Barcelo}).

As a specific example of the appearance of effective metrics due to
Lagrangian dynamics we mention that inviscid irrotational barotropic
hydrodynamics naturally falls into this scheme (which is why, with
hindsight, the derivation of the acoustic metric was so relatively
straightforward)~\cite{Unruh,Visser,Barcelo}. In inviscid irrotational
barotropic hydrodynamics the lack of viscosity (dissipation)
guarantees the existence of a Lagrangian; which a priori could depend
on several fields. Since the flow is irrotational $\vec v =
\nabla\vartheta$ is a function only of the velocity potential, and the
Lagrangian is a function only of this potential and the
density. Finally the equation of state can be used to eliminate the
density leading to a Lagrangian that is a function only of the single
field $\vartheta$ and its derivatives.

Note that in all these cases the (fundamental) dimensionality of
spacetime is put in by hand --- in the present formalism there is no
way to determine the fundamental dimensionality dynamically. (Of
course in a Kaluza--Klein framework the {\emph{effective}}
dimensionality can change if some dimensions become small for
dynamical reasons.) Also note that $d=1$ space dimensions is special,
and the present formulation does not work unless
$\det(f^{\mu\nu})=1$. This observation can be traced back to the
conformal covariance of the Laplacian in $1+1$ dimensions, and implies
(perhaps ironically) that the only time the procedure risks failure is
when considering a field theory defined on the world sheet of a
string-like object.

We next demonstrate that even if you do not have a Lagrangian, it is
still possible to extract an ``effective metric'' for a system with
one degree of freedom. (More precisely, we can define a conformal
class of effective metrics.  We shall see that the analysis is not as
geometrically ``clean''.)

\section{Second-order hyperbolic PDE:
Linearization and geometrical interpretation}

Consider an arbitrary second-order hyperbolic PDE written in the form
\begin{equation}
F(x, \phi, \partial_\mu \phi, \partial_\mu \partial_\nu \phi) = 0.
\end{equation}
The function $F$ and the field $\phi$ are taken to be real.  The PDE
does not have to be linear or even quasi-linear.  Defining
hyperbolicity for such a general equation is not trivial --- not even
Courant and Hilbert~\cite{Courant} deal with this particular case
explicitly. Instead we shall slightly adapt the definitions of Courant
and Hilbert, using them to define hyperbolicity for this system in
terms of the linearized equation.

So, suppose we linearize around some solution $\phi_0$, writing
\begin{equation}
\phi(t,\vec x) = 
\phi_0(t,\vec x) + 
\epsilon  \phi_1(t,\vec x) + O(\epsilon^2).
\end{equation}
Then
\begin{equation}
{\partial F\over\partial(\partial_\mu \partial_\nu \phi)} \;
 \partial_\mu \partial_\nu \phi_1 
+
{\partial F\over\partial(\partial_\mu \phi)} \;
 \partial_\mu \phi_1 
+
{\partial F\over\partial\phi} \;
 \phi_1 = 0.
\label{pdeone}
\end{equation}
Thus the fluctuation satisfies a second-order linear PDE with
time-dependent and position-dependent coefficients (these coefficients
again depend on the background field about which one is linearizing),
though the linear PDE looks somewhat different from what we
encountered in the Lagrangian analysis.  The linearized PDE is said to
be hyperbolic if the matrix ${\partial F/\partial(\partial_\mu
\partial_\nu \phi)}$ has ``Lorentzian'' signature
$\pm[-,+^d]$. This corresponds to standard mathematical usage, as
given for instance in volume 2 of Courant and Hilbert~\cite{Courant}
or in the Encyclopedic Dictionary of Mathematics~\cite{EDM}.

To give a geometrical interpretation to the linearized PDE we start by
regrouping the coefficients as follows:
\begin{equation}
\partial_\mu \left\{ 
{\partial F\over\partial(\partial_\mu \partial_\nu \phi)} \;
\partial_\nu \phi_1 
\right\}
+
\left\{ 
{\partial F\over\partial(\partial_\mu \phi)} -  
\partial_\mu \left(
{\partial F\over\partial(\partial_\mu \partial_\nu \phi)}
\right)
\right\}\;
\partial_\mu \phi_1 
+
{\partial F\over\partial\phi} \;
 \phi_1 = 0.
\label{E:pde2}
\end{equation}
Now define
\begin{equation}
f^{\mu\nu} \equiv  {\partial F\over\partial(\partial_\mu \partial_\nu \phi)},
\end{equation}
and
\begin{equation}
\Gamma^{\mu} \equiv {\partial F\over\partial(\partial_\mu \phi)} -  
\partial_\mu \left(
{\partial F\over\partial(\partial_\mu \partial_\nu \phi)}
\right).
\end{equation}
Then
\begin{equation}
\partial_\mu \left\{ f^{\mu\nu} \partial_\nu \phi_1 \right\}
+
\Gamma^\mu \partial_\mu \phi_1 
+
{\partial F\over\partial\phi}  \; \phi_1 = 0.
\label{E:pde3}
\end{equation}
To complete the geometrical interpretation, it is most elegant to
modify the formalism developed for Lagrangian dynamics and define a
conformal class of metrics by
\begin{equation}
\Omega^{-(d-1)} \; \sqrt{-g}\; g^{\mu\nu}(\phi) \equiv f^{\mu\nu}.
\end{equation}
Here $\Omega$ is (for now) a free variable, which will be chosen to
simplify the final result. In terms of this metric the PDE becomes
\begin{equation}
\Omega^{-(d-1)} \;\sqrt{-g} \; \Delta \phi_1
+
\left[ \Gamma^\mu - (d-1)  \Omega^{-(d-1)} \; 
\sqrt{-g}\; g^{\mu\nu} \partial_\nu \ln\Omega \right] 
\partial_\mu \phi_1 
+
{\partial F\over\partial\phi}  \; \phi_1 = 0.
\label{E:pde4}
\end{equation}
We now construct the vector
\begin{equation}
A^\mu(\phi_0)  
=
{\Omega^{(d-1)}\over \sqrt{-g}} \;  
\left[
\Gamma^\mu - (d-1)  \Omega^{-(d-1)} \; \sqrt{-g}\; g^{\mu\nu} \; 
\partial_\nu \ln\Omega 
\right] 
= {\Omega^{(d-1)}\over \sqrt{-g}} \;
\left[
\Gamma^\mu - (d-1)  f^{\mu\nu} \; \partial_\nu \ln\Omega 
\right] 
\end{equation}
and define the potential
\begin{equation}
V(\phi_0) = - {\Omega^{(d-1)}\over\sqrt{-g}}\; {\partial F\over\partial\phi}.
\end{equation}
After all this the linearized PDE becomes:
\begin{equation}
\left[ 
\,\Delta(g(\phi_0)) + A^\mu(\phi_0) \; \partial_\mu - V(\phi_0) \,
\right] \phi_1 = 0.
\end{equation}
This is the geometrical formulation we are searching for.

The conformal factor $\Omega$, buried in the definitions of
$g_{\mu\nu}$, $A^\mu$, and $V$, is at this stage arbitrary and may be
chosen to simplify the formul\ae{} below according to some convenient
prescription.  In particular, and in contrast to the Lagrangian-based
analysis, the linearized PDE is not automatically self-adjoint. Formal
self-adjointness is equivalent to being able to choose
$A_\mu(\phi_0)=0$.  This can be done if and only if the covariant
vector
\begin{equation}
B_\mu = [f^{-1}]_{\mu\nu} \; \Gamma^\nu =
\left[ 
\left({\partial F\over\partial(\partial_\bullet \partial_\bullet \phi)} 
\right)^{-1}\right]_{\mu\nu} \;
\left\{
{\partial F\over\partial(\partial_\nu \phi)}
- \partial_\sigma \left(
{\partial F\over\partial(\partial_\sigma \partial_\nu \phi)}
\right)
\right\}
\end{equation}
is exact. Observe that this vector is defined directly in terms of the
coefficients in the linearized PDE, without having to pre-choose the
conformal factor $\Omega(\phi)$. If $B_\mu$ is exact ($B_\mu =
\partial_\mu\Theta$) then we can choose $\Omega$ according to 
the prescription
\begin{equation}
\Omega^{d-1} =\exp(\Theta).
\end{equation}
This has the effect of fixing the conformal factor $\Omega$ and the
metric $g$ in such a way that $A^\mu=0$. In this case the linearized
PDE is self-adjoint and extremely compact
\begin{equation}
\left[\Delta(g(\phi_0)) - V(\phi_0)\right] \phi_1 = 0,
\end{equation}
where $\Delta$ is the d'Alembertian operator associated with the
effective metric $g$, and $V$ is the scalar potential. This finally is
identical in form to the equation derived on the basis of the
Lagrangian analysis.

While several of the technical details are different from the
Lagrangian-based analysis, the basic flavor is the same: The key point
is that hyperbolicity of the linearized PDE is defined in terms of the
presence of a matrix of indefinite signature $\pm[-,+^d]$. This matrix
is enough to define a conformal class of Lorentzian metrics, and
picking the ``right'' member of the conformal class is largely a
matter of taste --- do whatever makes the ``geometrized'' equation
look cleanest. (In particular if $B_\mu$ is exact there is a unique
conformal class that makes the linearized PDE self-adjoint.)

It is very important to stress that the hyperbolic character of these
systems (versus elliptic, parabolic, or ``other'') is encoded in the
coefficients of the second-derivative terms in (\ref{pdeone}) [or
(\ref{laeone})]. If we assume a hyperbolic equation, then the null
cones defined by the metric $g_{\mu\nu}(x)$ contain all the pointwise
information about the propagation of waves at arbitrarily large
momentum. Equivalently, the null cones determine the propagation of
sharp pulses. In fact since the characteristic surfaces, defined by
\begin{equation}
f(t,\vec x) = 0, 
\qquad \hbox{with} \qquad
g^{\mu\nu} \; \partial_{\mu}f \; \partial_{\nu} f = 0,
\end{equation}
are independent of an overall conformal factor in
$g^{\mu\nu}\propto{\partial F/\partial(\partial_\mu
\partial_\nu \phi(q))}$ they take into account only the causal
structure of the geometry.  Wave propagation at low momenta (as
opposed to eikonal ray propagation at high momenta) will depend in
addition on the conformal factor.  That is, additional features of
wave propagation will show up when looking at low momenta --- where
the linear and zero-order terms in the PDE will be crucial.

\section{Induced gravity}

At this stage we have derived the existence of a classical background
metric $g_{\mu\nu}(\phi_0)$ and linearized fluctuations governed by
the equation
\begin{equation}
\left[\Delta(g(\phi_0)) - V(\phi_0)\right] \phi_1 = 0,
\end{equation}
To quantize these linearized fluctuations, and see the manner in which
they can back-react on the geometry, we adopt the standard one loop
background field formalism. That is, we go all the way back to the
classical action given in equation (\ref{E:bare-action}) and write the
fundamental field $\phi$ as the sum of a background field (not
necessarily satisfying the classical equations of motion) plus a
quantum fluctuation
\begin{equation}
\phi = \phi_\background + \phi_\quantum.
\end{equation}
Then it is a standard result that integrating out the quantum
fluctuations leads to the one-loop result
\begin{equation}
\Gamma[\phi_\background] = S[\phi_\background] + 
{1\over2} \;\hbar \; {\rm tr}\ln \left[
\left.{\delta^2 S\over\delta\phi\;\delta\phi} \right|_{\phi_\background}
\right]
+ O(\hbar^2).
\end{equation}
But in view of our classical analysis performed in section
(\ref{S:lagrangian}) we know that
\begin{equation}
\left.{\delta^2 S\over\delta\phi\;\delta\phi} \right|_{\phi_\background}
=
\Delta(g(\phi_\background)) - V(\phi_\background).
\end{equation}
So we can write
\begin{equation}
\Gamma[g(\phi_\background),\phi_\background] = S[\phi_\background] + 
{1\over2} \;\hbar \; {\rm tr}\ln \left[
\Delta(g(\phi_\background)) - V(\phi_\background)
\right]
+ O(\hbar^2).
\end{equation}
Here the determinant of the differential operator may be defined in
terms of zeta functions or heat kernel
expansions~\cite{Dowker,Hawking,Blau,Birrell-Davies}. Note that we
have chosen the notation to emphasize the fact that the effective
action depends on the background field in two ways:
{\emph{explicitly}} through $\phi_\background$, and
{\emph{implicitly}} through $g(\phi_\background)$. The key point (more
or less equivalent to Sakharov's ``induced gravity'' proposal) is that
defining the determinant requires both regularization and
renormalization, and that doing so introduces counterterms
proportional to the first $d/2$ Seeley-DeWitt
coefficients~\cite{DeWitt,Bunch,Blau,Birrell-Davies}. The form of
these counterterms is well-known, and in fact for the second-order
differential operator $ \Delta(g(\phi_\background)) -
V(\phi_\background)$ we have
\begin{eqnarray}
a_0 &=& 1;
\\
a_1 &=& {1\over 6} R(g) - V(\phi_\background);
\\
a_2 &=& 
{1\over2} \left({1\over 6} R - V(\phi_\background) \right)^2 
+ 
{1\over6}  \Delta V(\phi_\background)
-
{1\over30} \Delta R
-
{1\over180} R^{\mu\nu}\; R_{\mu\nu}
+
{1\over180} R^{\mu\nu\rho\sigma}\; R_{\mu\nu\rho\sigma}.
\end{eqnarray}
The higher-order Seeley--DeWitt coefficients are multinomials in the
Riemann tensor, its contractions, and covariant derivatives, and in
the potential $V(\phi_\background)$ and its covariant derivatives.
The zeroth Seeley--DeWitt coefficient $a_0$ induces a cosmological
constant, while $a_1$ induces an Einstein--Hilbert term. There are
additional terms proportional to $a_2$ [and $a_3$ and higher in more
than (3+1) dimensions].  The current experimental constraints on these
terms are rather weak~\cite{a2-limit}.

In the original version of Sakharov's idea~\cite{Sakharov}, he
introduced an explicit ultraviolet cutoff --- this approach is
equivalent to writing
\begin{equation}
\ln \left[\Delta(g(\phi_\background)) - 
V(\phi_\background)\right]_{\mathrm{cutoff}}
=
\int_{\mathrm{cutoff}}^\infty {ds\over s} 
\exp( - s  \left[\Delta(g(\phi_\background)) - V(\phi_\background)\right]).
\end{equation}
The asymptotic expansion of the heat kernel for small $s$ then reads
\begin{equation}
\langle x |
\exp( - s  \left[\Delta(g(\phi_\background)) - V(\phi_\background)\right])
| x \rangle =
{1\over(4\pi s)^{d/2}} \left[ \sum_{n=0}^N a_n \; s^n + O(s^{N+1}) \right]
\end{equation}
So that
\begin{equation}
\langle x | 
\ln \left[
\Delta(g(\phi_\background)) - V(\phi_\background)
\right]_{\mathrm{cutoff}} 
| x \rangle
=
{1\over(4\pi)^{d/2}} 
\left[ \sum_{n=0}^{d/2} a_n \; (\hbox{cutoff})^{n-d/2} + 
\hbox{finite} \right].
\end{equation}
As usual, one should interpret $(\hbox{cutoff})^0$ as
$\ln(\hbox{cutoff})$~\cite{Zuber}. Again we see that the effective
action contains terms proportional to a cosmological constant, the
Einstein--Hilbert action, and others. It is most useful to organize
the terms in the effective action in a gradient expansion in the
effective metric and background field. All in all:
\begin{equation}
\Gamma[g(\phi_\background),\phi_\background] = S[\phi_\background] +
\hbar \int \sqrt{-g}\; \kappa 
\left[ -2 \Lambda + R(g(\phi_\background)) \right] \d^{d+1}x   
+ \hbar X[g(\phi_\background),\phi_\background]
+ O(\hbar^2).
\end{equation}
Here $\kappa$ and $\Lambda$ are constants of dimensions $(mass)^2$
[that is, $(length)^{-2}$] that emerge from the renormalization
procedure.  Finally $X(\phi_\background)$ denotes all other finite
contributions to the renormalized one-loop effective action (including
gradient and curvature-squared contributions coming from $a_2$, plus
possible background-field dependent modifications of $\kappa$ and
$\Lambda$ coming from $a_1$ and $a_2$).  Phenomenologically, we will
want to eventually relate $\kappa$ to the Newton constant, and
$\Lambda$ to the cosmological constant.  It is the automatic emergence
of the Einstein--Hilbert action as part of the one-loop effective
action that is the salient point.

Note that our approach is not identical to Sakharov's idea --- in his
proposal the metric was put in by fiat, but without any intrinsic
dynamics; all the dynamics was generated via one loop quantum
effects. (Implicit in Sakharov's approach is the assumption that if
there is high energy microphysics leading to the notion of the metric,
then it should decouple from the low-energy physics; more on this
point below.) In our proposal the very existence of the effective
metric itself is an emergent phenomenon. In Sakharov's approach the
metric was free to be varied at will, leading precisely to the
Einstein equations (plus quantum corrections); in our approach the
metric is not a free variable and the equations of motion will be a
little trickier.

The quantum equations of motion are defined in the usual way by
varying with respect to the background $\phi_\background$. It is
important to remember that the metric is a function of the background
field so that it does not make sense to vary the metric independently
--- we must always evaluate variations using the chain rule. Thus
\begin{equation}
{\delta\Gamma[g(\phi_\background),\phi_\background]
\over
\delta\phi_\background(x)} 
\equiv
\left.{\delta\Gamma[g(\phi_\background),\phi_\background]
\over
\delta\phi_\background(x)}\right|_{g_\background}  +
\left.{\delta\Gamma[g(\phi_\background),\phi_\background]
\over\delta g_{\mu\nu}}\right|_{\phi_\background}
{\delta g_{\mu\nu}(\phi_\background)\over\delta\phi_\background(x)}.
\end{equation}
Applied to the one-loop action the equations of motion are
\begin{eqnarray}
\label{E:eom1}
{\delta\Gamma[g(\phi_\background),\phi_\background]
\over
\delta\phi_\background(x)} 
= &0& = 
\left\{
{\delta S[\phi_\background]\over\delta \phi_\background } 
+ 
\hbar
\left.
{\delta X[g(\phi_\background),\phi_\background]\over\delta \phi_\background }
\right|_{g_\background} 
\right\}
\nonumber
\\
&&
+
\hbar \left\{
\kappa(\phi_\background) \sqrt{g} 
\left[ G^{\mu\nu}(g) + \Lambda(\phi_\background) g^{\mu\nu} \right]  \;
+
\left.
{\delta X[g(\phi_\background),\phi_\background]\over\delta g_{\mu\nu} }
\right|_{\phi_0} 
\right\}
{\delta g_{\mu\nu}(\phi_\background)\over\delta\phi_\background(x)} +
O(\hbar^2).
\end{eqnarray}
These are not the Einstein equations, but they are closely related:
the first two terms on the RHS are problematic in that they encode the
dependence on the original background geometry (typically Minkowski
space) implicitly used in setting up the fundamental Lagrangian, while
the next three terms involve the effective metric and are similar to
those obtained in studies of embedded-submanifold versions of general
relativity; see for example Regge and Teitelboim~\cite{Regge} and
Deser {\etal}~\cite{Deser}.  To obtain Einstein-like dynamics we need
to suppress the dependence on the original background metric, and
write these equations solely in terms of the effective metric
$g(\phi_0)$. This can be done if and only if there exists a functional
$ Z[g(\phi_0)]$, which depends on $\phi_0$ only implicitly via the
effective metric, such that
\begin{equation}
\label{E:decouple}
\int 
\L[\phi_\background] 
\; \d^{d+1}x
+ \hbar \; X[g(\phi_\background),\phi_\background] 
= \hbar \; Z[g(\phi_\background)] + O(\hbar^2).
\end{equation}
This is, very definitely, a powerful restriction of the theory under
consideration; but it appears to be the minimal condition for
something similar to Einstein dynamics to arise.  In particular we
view this particular hurdle as the single biggest issue facing the
``induced gravity'' proposal, though some particular implementations
of this idea may skirt the issue. For instance, consider Volovik's
proposals to extract the dynamics of Einstein gravity from condensed
matter quasiparticle excitations~\cite{Volovik-disp}: Volovik
essentially argues that certain theories might exhibit ``one-loop
dominance'' in the sense that the one-loop physics dominates over the
zero loop physics. In contrast, Sakharov implicitly assumes that
whatever the microphysics is, it has effectively decoupled from the
low energy effective metric. Our expression in equation
(\ref{E:decouple}) above is an explicit characterization of what this
decoupling condition should be.  Note that it is quite sufficient for
our purposes if this constraint holds as an approximation for some
region in field space surrounding the metrics of interest; it does not
need to be a global constraint on the theory. Given this constraint
the background geometry (the microphysics) decouples from the
effective metric (the macrophysics) and we have
\begin{equation}
\left\{ 
\kappa \left[G^{\mu\nu}(g) + \Lambda \; g^{\mu\nu}\right] + 
{1\over\sqrt{g}}{\delta Z[g(\phi_\background)]\over\delta g_{\mu\nu} }
\right\} \; 
{\delta g_{\mu\nu}(\phi_\background)\over\delta\phi_\background(x)} = O(\hbar).
\end{equation}
The ${\delta Z[g]/\delta g}$ term now encodes three separate terms
form equation (\ref{E:eom1}) and denotes the type of ``curvature
squared'' corrections to the Einstein equations that are commonly
encountered in string theory (indeed in almost any candidate theory
for quantum gravity) and in the usual implementation of Sakharov's
approach.  Additionally, it must be emphasised that because of the
contraction with the ${\delta g_{\mu\nu}(\phi_\background)/
\delta\phi_\background(x)}$ these are not the usual Einstein
equations, though they are certainly implied by the (curvature
enhanced) Einstein equations. It is in this sense that we can begin to
see the structure of Einstein gravity emerging from this
field-theoretic normal mode analysis.

\section{Discussion}

In this paper we have provided two key developments:
\begin{enumerate}
\item 
We have shown that the emergence of an ``effective metric'', in the
sense that this notion is used in the so-called ``analog models'' of
general relativity, is a rather generic feature of the linearization
process. While the existence of an effective metric by itself does not
allow you to simulate all of Einstein gravity, it allows one to do
quite enough to be really significant --- in particular it seems that
the existence of an effective Lorentzian metric is really all that is
in principle needed to obtain simulations of the Hawking radiation
effect~\cite{Unruh,Hawking-radiation,without-entropy}. In this regard,
the major technical limitation of the current analysis is that it is
limited to a single scalar field; extensions of this idea involve both
some technical subtleties and some new physics, and we shall discuss
that scenario more fully elsewhere. The major piece of additional
physics is the possible presence of birefringence, or more generally
``multi-refringence'', with different normal modes possibly reacting
to different metrics. The {\Eotvos} experiment [the observational
universality of free fall to extremely high accuracy] indicates that
all the physical fields coupling to ordinary bulk matter ``see'' to
high precision the {\emph{same}} metric, allowing us to formulate the
Einstein Equivalence principle and speak of {\emph{the}} metric of
spacetime. Thus in extending the notion of effective geometry to a
system with many degrees of freedom, experiment tells us that we
should seek conditions that would naturally serve to suppress
birefringence.  Only in that case would it make sense to speak of a
unique spacetime metric (or at worst, of multiple almost-degenerate
metrics).
\item
By invoking one-loop quantum effects, we can argue that something akin
to Sakharov's induced gravity scenario is operative: in particular we
can generically argue that there is a term in the quantum effective
action proportional to the Einstein--Hilbert action. However because
of the technical assumption that the effective metric depends on the
background only via the single scalar field $\phi_0(x)$ we have not
been able to reproduce full Einstein gravity, though certainly have
some extremely suggestive results along this line.  Additional issues
that are definitely worth future exploration are the physical import
of the fine tuning used to decouple the effective metric from the
background [equation (\ref{E:decouple})], the question of going beyond
the linearized approximation [that is, beyond one loop], and whether
the addition of extra fields helps one to obtain a better
approximation to full Einstein gravity --- this because you would get
one equation of motion per background field, so with six or more
fields you would expect to be able to explore the full algebraic
structure of the metric. So adding extra fields, which is technically
a hindrance in the kinematical part of the program (developing the
effective metric formalism), should in compensation allow one to more
closely approach the dynamics of Einstein gravity.
\end{enumerate}

In summary: The full generality of the situations under which
effective metrics are encountered is truly remarkable, and the extent
to which the resulting analog models seem able to reproduce key
aspects of Einstein gravity is even more remarkable. The physics of
these systems is fascinating, and the potential for laboratory
investigation of models close to (but not necessarily identical to)
Einstein gravity is extremely encouraging.  Our interpretation of
these results is that they provide suggestive evidence that what we
call Einstein gravity (general relativity) is an almost automatic
low-energy consequence of almost any well behaved quantum field
theory: the dimensionality of spacetime is put in by hand, but the
occurrence of an effective metric is almost automatic (even in the
classical theory), while the presence of Einstein-like dynamics can
plausibly be engendered by one-loop quantum effects.

\section*{Acknowledgements}

The research of Matt Visser was supported by the US Department of
Energy.  Stefano Liberati was supported by the US National Science
Foundation. Carlos {\Barcelo} was supported by the Spanish Ministry of
Education and Culture (MEC).  The authors are indebted to Grigori
Volovik for the clear presentation of his ideas on induced gravity in
superfluid quasiparticle systems as developed in his lectures at the
Workshop on ``Analog Models of General Relativity''~\cite{Workshop}.
The authors are particularly grateful to Sebastiano Sonego for
extensive discussions and input; and also wish to thank Ted Jacobson
for his interest and comments.


\end{document}